\documentclass[aps,twocolumn,prx,preprintnumbers,showpacs,amsmath,amssymb,superscriptaddress]{revtex4-1}
\usepackage[pdftex, colorlinks, citecolor=blue]{hyperref}   
\usepackage{graphicx}
\usepackage{dcolumn}
\usepackage{threeparttable}
\usepackage{multirow}
\usepackage{booktabs}
\usepackage{txfonts}
\usepackage{xcolor}
\usepackage{bm}
\usepackage{amssymb}
\usepackage{amsmath}
\usepackage{latexsym}
\usepackage{epsfig}
\usepackage{amsbsy}
\usepackage{array}
\usepackage{tabularx}
\usepackage{esvect} 
\usepackage{extarrows}
\usepackage{float}
\usepackage[british]{babel}

\begin{document}

\title{Symmetry breaking induced insulating electronic state in Pb$_{9}$Cu(PO$_4$)$_6$O}

\author{Jiaxi Liu}
\thanks{These authors contribute equally to this work.}
\affiliation{%
Shenyang National Laboratory for Materials Science, Institute of Metal Research, Chinese Academy of Sciences, 110016 Shenyang, China.
}%
\affiliation{%
School of Materials Science and Engineering, University of Science and Technology of China, 110016 Shenyang, China.
}%

\author{Tianye Yu}
\thanks{These authors contribute equally to this work.}
\affiliation{%
Shenyang National Laboratory for Materials Science, Institute of Metal Research, Chinese Academy of Sciences, 110016 Shenyang, China.
}%

\author{Jiangxu Li}
\email{jxli15s@imr.ac.cn}
\affiliation{%
Shenyang National Laboratory for Materials Science, Institute of Metal Research, Chinese Academy of Sciences, 110016 Shenyang, China.
}%

\author{Jiantao Wang}
\affiliation{%
Shenyang National Laboratory for Materials Science, Institute of Metal Research, Chinese Academy of Sciences, 110016 Shenyang, China.
}%
\affiliation{%
School of Materials Science and Engineering, University of Science and Technology of China, 110016 Shenyang, China.
}%

\author{Junwen Lai}
\affiliation{%
Shenyang National Laboratory for Materials Science, Institute of Metal Research, Chinese Academy of Sciences, 110016 Shenyang, China.
}%
\affiliation{%
School of Materials Science and Engineering, University of Science and Technology of China, 110016 Shenyang, China.
}%

\author{Yan Sun}
\affiliation{%
Shenyang National Laboratory for Materials Science, Institute of Metal Research, Chinese Academy of Sciences, 110016 Shenyang, China.
}%

\author{Xing-Qiu Chen}%
\affiliation{%
Shenyang National Laboratory for Materials Science, Institute of Metal Research, Chinese Academy of Sciences, 110016 Shenyang, China.
}%

\author{Peitao Liu}%
\email{ptliu@imr.ac.cn}
\affiliation{%
Shenyang National Laboratory for Materials Science, Institute of Metal Research, Chinese Academy of Sciences, 110016 Shenyang, China.
}%

\begin{abstract}
The recent experimental claim of room-temperature ambient-pressure superconductivity in a Cu-doped lead-apatite (LK-99)
has ignited substantial research interest in both experimental and theoretical domains.
Previous density functional theory (DFT) calculations with the inclusion of an on-site Hubbard interaction $U$
consistently predict the presence of flat bands crossing the Fermi level.
This is in contrast to DFT plus dynamical mean field theory calculations, which reveal
the Mott insulating behavior for the stoichiometric Pb$_{9}$Cu(PO$_4$)$_6$O compound.
Nevertheless, the existing calculations are all based on the $P6_3/m$ structure,
which is argued to be not the ground-state structure.
Here, we revisit the electronic structure of Pb$_{9}$Cu(PO$_4$)$_6$O
with the energetically more favorable $P\bar{3}$ structure, fully taking into account electronic symmetry breaking.
We examine all possible configurations for Cu substituting the Pb sites.
Our results show that the doped Cu atoms exhibit a preference for substituting the Pb2 sites than the Pb1 sites.
In both cases, the calculated substitutional formation energies are large, indicating the difficulty in incorporating Cu at the Pb sites.
We find that most of structures with Cu at the Pb2 site tend to be insulating,
while the structures with both two Cu atoms at the Pb1 sites (except one configuration)
are predicted to be metallic by DFT+$U$ calculations.
However, when accounting for the electronic symmetry breaking,
some Cu-doped configurations previously predicted to be metallic
(including the structure studied in previous DFT+$U$ calculations) become insulating.
Our work highlights the importance of symmetry breaking in obtaining
correct electronic state for Pb$_{9}$Cu(PO$_4$)$_6$O, thereby reconciling previous DFT+$U$ and DFT+DMFT calculations.
\end{abstract}

\maketitle

\section{Introduction}

Due to the potential for technological revolutions, the search for high-$T_c$ superconductors
under ambient conditions has persisted as a longstanding aspiration for both experimental and theoretical scientists.
In light of this, the recent report claiming room-temperature ambient-pressure superconductivity and diamagnetism
in a Cu-doped lead-apatite (LK-99)~\cite{Lee2023_1,Lee2023_2} has sparked substantial excitement not only
within the physics community but also in the fields of chemistry and materials science.
Several independent experimental groups have tried to synthesize
LK-99~\cite{arXiv:2307.16802,arXiv:2307.16402,arXiv:2308.01516,arXiv:2308.01723,arXiv:2308.03110,arXiv:2308.03544v3,
arXiv:2308.03823,arXiv:2308.04353,arXiv:2308.05222,arXiv:2308.07800, arXiv:2308.05001,arXiv:2308.05776,
wen_arXiv:2303.08759,arXiv:2308.06256,arXiv:2308.05778,arXiv:2308.06589},
but the claimed superconductivity of LK-99 has not been obtained.
Notably, the experimentally observed sudden decrease in resistivity at approximately 400 K
was argued to be attributed to a first-order transition from the high-temperature insulating $\beta$-Cu$_2$S phase
to the low-temperature semiconducting $\gamma$-Cu$_2$S phase~\cite{arXiv:2308.04353,arXiv:2308.05222,arXiv:2308.07800},
and the half levitation observed in experiments was explained by the presence of soft ferromagnetism~\cite{arXiv:2308.03110}.
Moreover, insulating transport behaviors were observed
by several experimental groups in their synthesized samples~\cite{arXiv:2308.06589, arXiv:2307.16402, arXiv:2308.05778, wen_arXiv:2303.08759, arXiv:2308.06256}.
In contrast, density functional theory (DFT) calculations~\cite{JMST2023,arXiv:2307.16892,Liang1,arXiv:2308.00698,arXiv:2308.01135,arXiv:2308.01315,
arXiv:2308.02469,arXiv:2308.03218,OganovDMFT,Liang2,PhilippDMFT,arXiv:2308.05124,arXiv:2308.05143,arXiv:2308.05618,Liang3} consistently
predicted the presence of flat bands crossing the Fermi level due to the spatial separation of Cu atoms.
DFT plus dynamical mean field theory (DFT+DMFT) calculations~\cite{OganovDMFT,PhilippDMFT}, however, demonstrated that
the stoichiometric Pb$_{9}$Cu(PO$_4$)$_6$O compound is actually a Mott insulator due to the very large interaction versus bandwidth ratio $U/W>30$
and additional electron doping is needed to achieve a conducting state.
The calculations using the fluctuation exchange method showed that
spin and orbital fluctuations are too weak even under optimized conditions for superconductivity,
indicating the absence of superconductivity in Pb$_{9}$Cu(PO$_4$)$_6$O~\cite{Liang3}.

It is important to note that  the existing calculations are all based on the $P6_3/m$-Pb$_{10}$(PO$_4$)$_6$O structure model as refined by Krivovichev and Burns in 2003~\cite{ExptStr2003}. Very recently, Krivovichev~\cite{NEWstr2023} reinvestigated the crystal structure of Pb$_{10}$(PO$_4$)$_6$O by single-crystal x-ray diffraction (XRD) using crystals prepared by Merker and Wondratschek~\cite{MW1960} and concluded that the ground-state structure of Pb$_{10}$(PO$_4$)$_6$O is a superstructure with the $c$ lattice parameter doubled with respect to the $P6_3/m$ structure and exhibits a $P\bar{3}$ symmetry~\cite{NEWstr2023}.
According to the analysis of Krivovichev~\cite{NEWstr2023}, the $P6_3/m$-Pb$_{10}$(PO$_4$)$_6$O structure
may correspond to the Pb$_{10}$(PO$_4$)$_6$O$_x$(OH)$_{2-x}$  ($x$$\sim$0.4) member of the Pb$_{10}$(PO$_4$)$_6$O- Pb$_{10}$(PO$_4$)$_6$(OH)$_2$ solid solution series, or to the high-temperature polymorph of Pb$_{10}$(PO$_4$)$_6$O~\cite{NEWstr2023}.
This important finding therefore raises a critical question of
whether the conclusions drawn from previous calculations relying on the high-temperature $P6_3/m$-Pb$_{10}$(PO$_4$)$_6$O structure remain valid
when using the proposed ground-state $P\bar{3}$-Pb$_{10}$(PO$_4$)$_6$O structure.

\begin{figure*}
\begin{center}
\includegraphics[width=0.85\textwidth, clip]{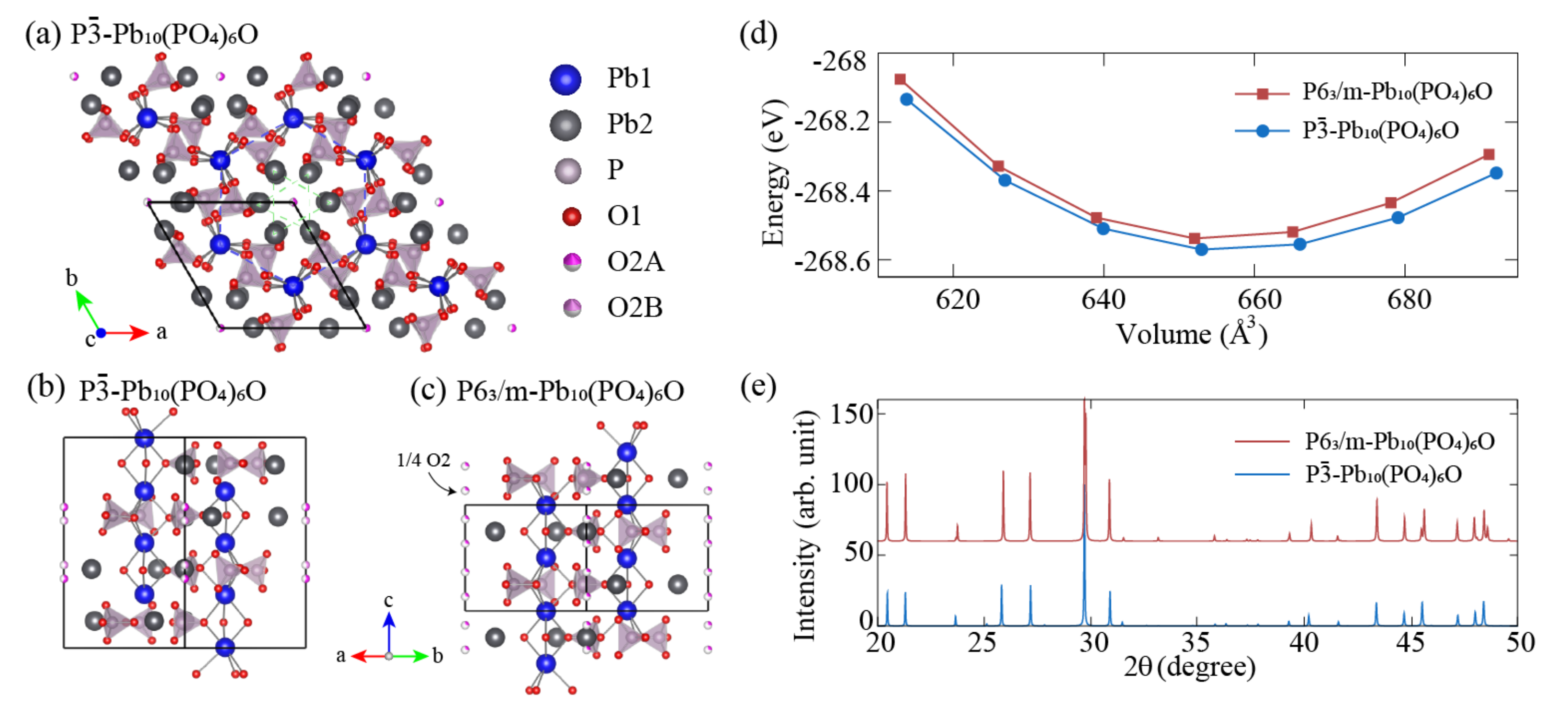}
\end{center}
\caption{(a) Top view of the crystal structure of $P\bar{3}$-Pb$_{10}$(PO$_4$)$_6$O.
(b) Side view of the crystal structure of $P\bar{3}$-Pb$_{10}$(PO$_4$)$_6$O.
(c) Side view of the crystal structure of $P6_3/m$-Pb$_{10}$(PO$_4$)$_6$O.
The black lines in (a)-(c) indicate the unit cell that is periodically replicated.
(d) Energy-volume curves. Note that for a direct comparison, the total energy and volume of the $P6_3/m$ structure are doubled.
(e) Simulated XRD patterns.
}
\label{Fig1_parent}
\end{figure*}

In this work, we revisit the electronic structure of Pb$_{9}$Cu(PO$_4$)$_6$O with the proposed $P\bar{3}$ ground-state structure
using the DFT+$U$ method and DFT+DMFT approach, fully accounting for electronic symmetry breaking (ESB).
We find that the newly refined $P\bar{3}$-Pb$_{10}$(PO$_4$)$_6$O superstructure is also an insulator,
but energetically more favorable than the $P6_3/m$ structure.
Both structures are found to be dynamically unstable due to the partial occupancy of ``extra" O atoms.
Regarding the Cu-doped case, we examine all possible configurations for Cu substituting the Pb sites,
leading to in total 44 symmetry-inequivalent configurations.
Our results show that the doped Cu atoms prefer to substitute the Pb2 sites
rather than the Pb1 sites that previous calculations assumed in their employed structures.
In both cases, the calculated substitutional formation energies are large.
The structures with Cu at the Pb2 sites tend to be insulating.
In contrast, the structures with both two Cu atoms at the Pb1 sites (except one configuration)
are predicted to be metallic by DFT+$U$.
However, the band gap is opened for some Cu-doped configurations
when the ESB is taken into account.
We note that our finding also applies to the $P6_3/m$-Pb$_{9}$Cu(PO$_4$)$_6$O structure studied in previous DFT+$U$ calculations.
This work stresses the importance of symmetry breaking in obtaining
correct insulating electronic state for Pb$_{9}$Cu(PO$_4$)$_6$O,
and therefore,  reconciles previous DFT+$U$ and DFT+DMFT calculations.

\section{Computational details}

 For all the DFT+$U$ calculations, we utilized computational settings similar to those employed in our previous work~\cite{JMST2023}.
Specifically, the Vienna \emph{ab initio} simulation package~\cite{PhysRevB.47.558, PhysRevB.54.11169} was used.
The plane-wave cutoff was chosen to be 520 eV.
A $\Gamma$-centered $k$-point grid with a $k$-spacing of 0.03 $2\pi/\AA$ was used for structural relaxations
and the $k$-spacing was reduced to 0.02 $2\pi/\AA$ for more accurate total energy and electronic structure calculations,
The electronic interactions were described using the Perdew-Burke-Ernzerhof (PBE) functional~\cite{PhysRevLett.77.3865}.
The VASP recommended projector augmented wave pseudopotentials~\cite{PhysRevB.50.17953,PhysRevB.59.1758} were employed.
The tetrahedron method with Bl\"{o}chl corrections was used for density of states (DOSs) calculations,
whereas the Gaussian smearing method with a smearing width of 0.05 eV was used for other calculations.
The convergence criteria for the total energy and ionic forces were set to 10$^{-6}$ eV and 0.01 eV/$\AA$, respectively.
The phonon dispersions were calculated by finite displacements using the Phonopy code~\cite{Togo2015}.
For phonon calculations,
a $2\times2\times2$ supercell (656 and 328 atoms for $P\bar{3}$- and $P6_3/m$-Pb$_{10}$(PO$_4$)$_6$O structures, respectively) was used
and the $k$-point grid was reduced to the $\Gamma$ point only.
The static electronic correlation effects were described using the Dudarev's DFT+$U$ scheme~\cite{PhysRevB.57.1505,PhysRevMaterials.3.083802}
with the Hubbard interaction $U$= 4 eV~\cite{PhysRevB.73.195107} on the Cu-$3d$ shell.

To study the effect of dynamical electronic correlation effects, we also conducted DFT+DMFT calculations~\citep{kotliar2006electronic,haule2010dynamical}.
For the DFT part, the all-electron WIEN2K package~\citep{blaha2001wien2k} was employed.
The adopted atomic spheres $R_\text{MT}$ were 2.13, 2.35, 1.53 and 1.39 Bohr for Cu, Pb, P and O, respectively,
and the plane-wave cutoff $K_\text{max}$ was set to $R_\text{MT}K_\text{max} = 7.0$.
The impurity problem in the DMFT calculation was solved by the continuous time
quantum Monte Carlo method~\citep{werner2006continuous,haule2007quantum} at a temperature of 387 K.
The spectral functions along the real-frequency axis were obtained
by analytical continuation using the maximum entropy method~\citep{haule2010dynamical}.

\section{Structural, electronic, and dynamical properties of lead-apatite}

\begin{figure*}
\begin{center}
\includegraphics[width=0.6\textwidth, clip]{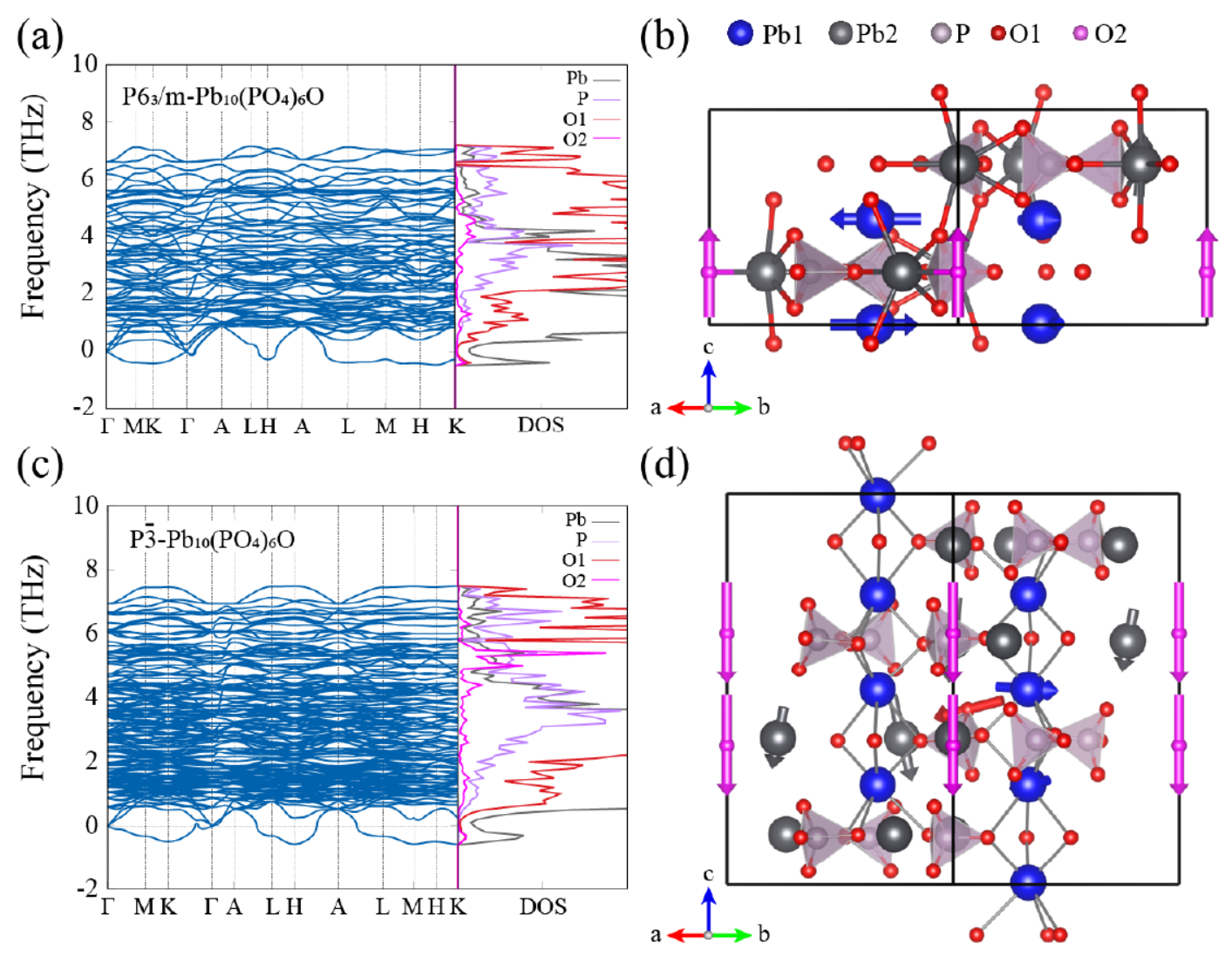}
\end{center}
\caption{Calculated phonon dispersions and DOSs, and atomic displacements (in arrows) associated with the imaginary frequency at M (0.5, 0, 0) point
for (a)-(b) $P6_3/m$-Pb$_{10}$(PO$_4$)$_6$O and (c)-(d) $P\bar{3}$-Pb$_{10}$(PO$_4$)$_6$O.
Note that for a better visualization of the phonon branches with imaginary frequencies, the phonon frequencies beyond 10 THz are not shown.
}
\label{Fig2_phonon}
\end{figure*}

Figures~\ref{Fig1_parent}(a)-(c) compare the crystal structures of $P\bar{3}$- and $P6_3/m$-Pb$_{10}$(PO$_4$)$_6$O.
In both structures, there exist two symmetry-inequivalent Pb atoms, named Pb1 and Pb2.
The Pb1 atoms form a hexagon, whereas the Pb2 atoms form two oppositely shaped triangles [see Fig.~\ref{Fig1_parent}(a)].
Along the $c$ axis, the Pb2 atoms along with the surrounding insulating PO$_4$ units form a cylindrical column centered at ``extra" O2 atoms [Fig.~\ref{Fig1_parent}(a)].
The noticeable difference between the two structures lies in the doubling of the $c$ lattice parameter in $P\bar{3}$-Pb$_{10}$(PO$_4$)$_6$O  [compare Figs.~\ref{Fig1_parent}(b) and (c)].
This is caused by the ordering of the O2 atoms within the structure channels along the $c$ axis~\cite{NEWstr2023}.
The O2 atoms form short bonds to the Pb2 atoms, leading to two different sites Pb2A and Pb2B.
We note that for the $P6_3/m$ structure, the O2 atoms are 1/4 occupied~\cite{ExptStr2003},
while for the $P\bar{3}$ structure, the O2A atoms are 0.55 occupied and O2B atoms are 0.45 occupied~\cite{NEWstr2023}.
To model these structures, we removed three out of four O2 atoms for the  $P6_3/m$ structure,
whereas for the $P\bar{3}$ structure only the O2B atoms were kept.
The calculated lattice parameters of both structures are compiled in Table~\ref{Table1},
showing good agreement with respective experimental values~\cite{ExptStr2003,NEWstr2023}
The overestimated predicted lattice parameters are due to the employed PBE functional.
The structural data of  $P\bar{3}$- and $P6_3/m$-Pb$_{10}$(PO$_4$)$_6$O are provided in the Supplementary Material~\cite{SM}.
The calculated energy-volume curves for both $P\bar{3}$- and $P6_3/m$-Pb$_{10}$(PO$_4$)$_6$O are shown Fig.~\ref{Fig1_parent}(d).
It is evident that the $P\bar{3}$ structure is more energetically favorable than the $P6_3/m$ structure across the entire range of volumes considered.
However, both structures display similar XRD patterns that are difficult to visually differentiate.
Upon closer examination, one can see that the $P\bar{3}$ structure exhibits two additional minor peaks at around 25 and 28 degrees.

\begin{table}
\caption{DFT predicted lattice parameters $a$ and $c$ (in $\AA$), and volume (in $\AA^3$)
of Pb$_{10}$(PO$_4$)$_6$O with $P6_3/m$ or $P\bar{3}$ symmetries.
The available experimental data are given for comparison.
}
\begin{ruledtabular}
\begin{tabular}{lccc}
   & $a$  &$c$ & Volume  \\
\hline
$P6_3/m$-Pb$_{10}$(PO$_4$)$_6$O (Calc.) & 10.024 & 7.482  & 651.02    \\
$P6_3/m$-Pb$_{10}$(PO$_4$)$_6$O (Expt.)\cite{ExptStr2003}  & 9.865 & 7.431  & 626.25   \\
$P\bar{3}$-Pb$_{10}$(PO$_4$)$_6$O (Calc.) & 10.019 &15.034  & 1306.80    \\
$P\bar{3}$-Pb$_{10}$(PO$_4$)$_6$O (Expt.)\cite{NEWstr2023}  &  9.811 & 14.840  & 1237.06  \\
\end{tabular}
\end{ruledtabular}
\label{Table1}
\end{table}

Furthermore, we conducted phonon calculations and found that both structures are dynamically unstable (see Fig.~\ref{Fig2_phonon}).
The calculated phonon density of states (DOSs) indicate that the imaginary phonon modes for both structure
arise from the partial occupancy of ``extra" O2 atoms and associated neighboring Pb atoms (Fig.~\ref{Fig2_phonon}).
Our calculated phonon instability of $P6_3/m$-Pb$_{10}$(PO$_4$)$_6$O is consistent with previous DFT results~\cite{arXiv:2308.05618,Xiayi}.
Turning to the electronic structure, we found that similar to the $P6_3/m$ compound, the $P\bar{3}$-Pb$_{10}$(PO$_4$)$_6$O compound
is also a nonmagnetic insulator, with a PBE-predicted indirect gap of 2.96 eV (Supplementary Material Fig.~S1~\cite{SM}).
The obtained insulating nature of lead-apatite is consistent with the experimental observation~\cite{Lee2023_2}.

\section{Electronic structure of copper-doped lead-apatite}

\begin{figure*}
\begin{center}
\includegraphics[width=0.8\textwidth, clip]{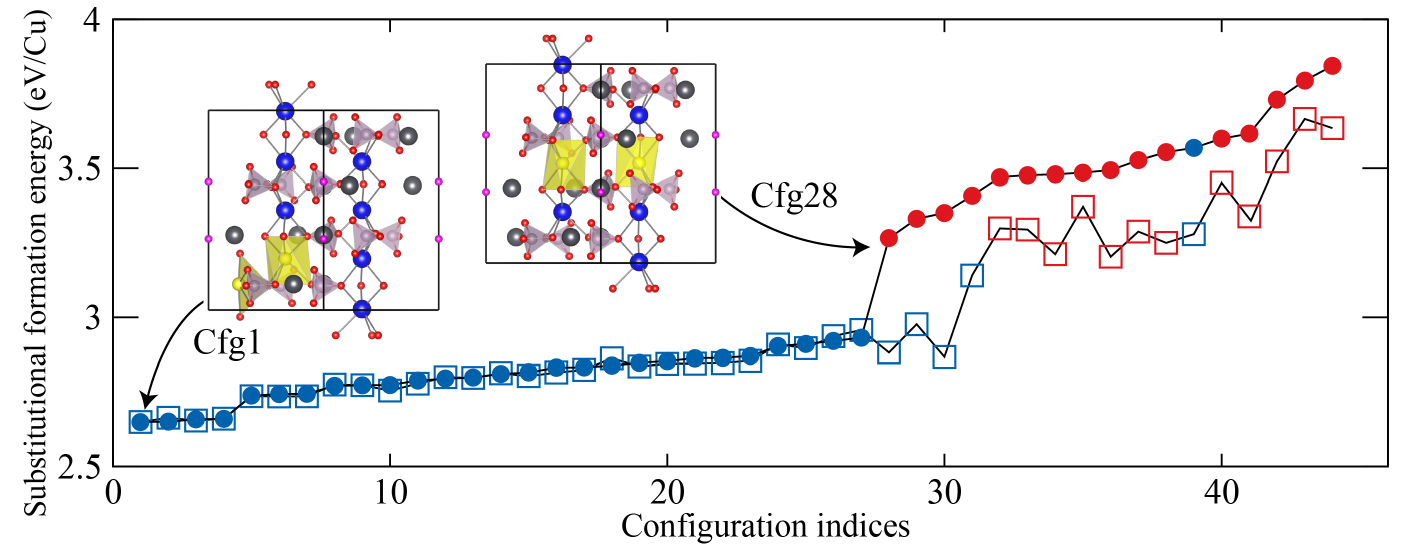}
\end{center}
\caption{Computed Cu$\rightarrow$Pb substitutional formation energies ($E_f$)  for forty-four symmetry-inequivalent configurations.
The configurations are arranged in ascending magnetically collinear DFT+$U$ (without ESB) calculated $E_f$ (in circles),
and their crystal structures are shown in Supplementary Material Fig.~S2~\cite{SM}.
The DFT+$U$+SOC calculated $E_f$ are displayed in squares.
The blue and red colors indicate the insulating and metallic states, respectively.
We note that the configuration Cfg39 is found to be a robust FM insulator,
regardless of ESB or variation of $U$.
All the data associated with this plot are provided in Supplementary Material Table~S1~\cite{SM}.
}
\label{Fig3_Ef}
\end{figure*}

\begin{figure*}
\begin{center}
\includegraphics[width=0.85\textwidth, clip]{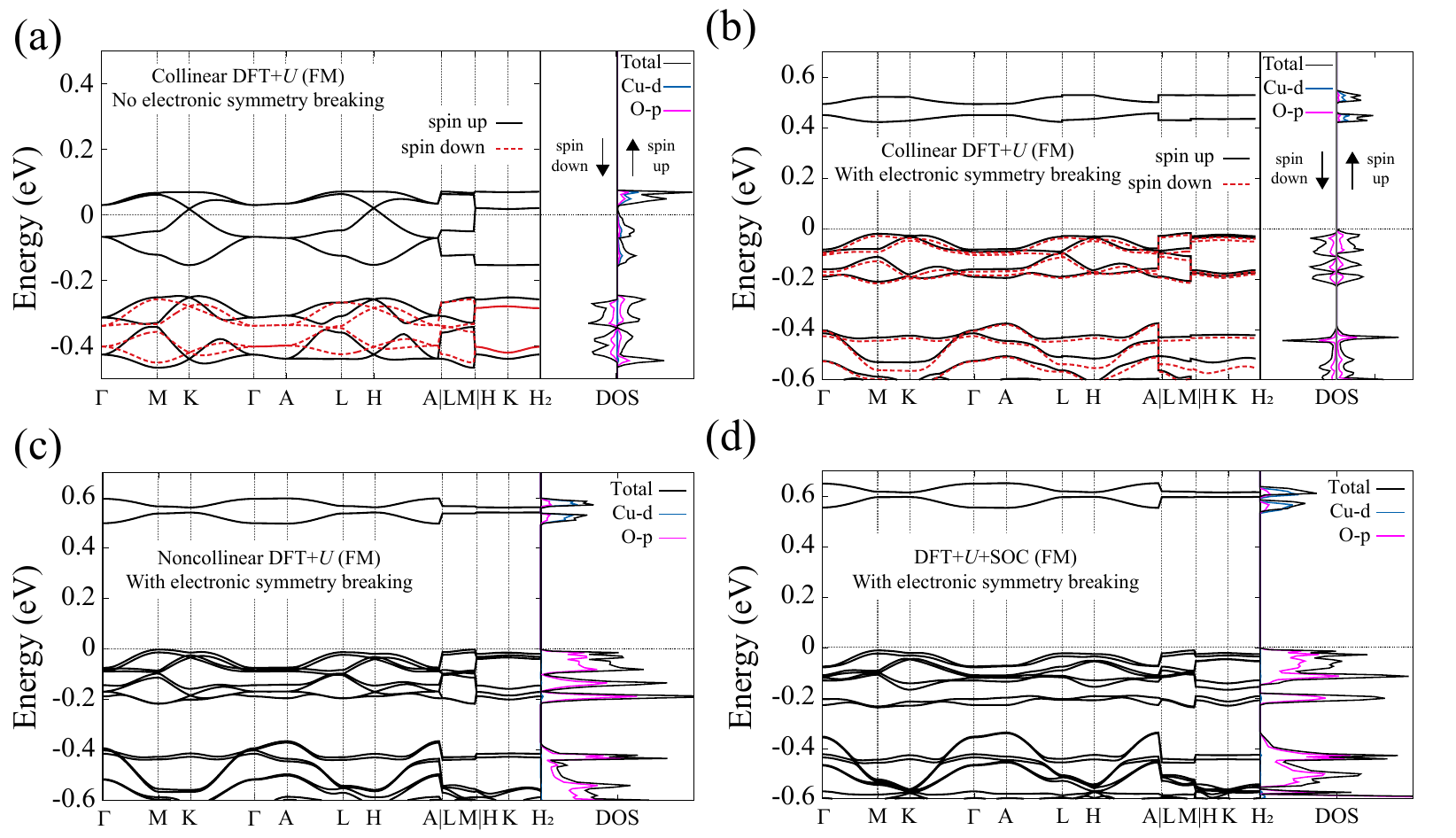}
\end{center}
\caption{Predicted electronic band structure and density of states for Cfg28-$P\bar{3}$-Pb$_{9}$Cu(PO$_4$)$_6$O
by (a) magnetically collinear DFT+$U$ without ESB,
(b) magnetically collinear DFT+$U$ with ESB,
(c) magnetically noncollinear DFT+$U$ with ESB,
and (d) DFT+$U$+SOC with ESB. A ferromagnetic setup is adopted for all the calculations.
}
\label{Fig4_all_bands}
\end{figure*}

Having identified the ground-state structure of Pb$_{10}$(PO$_4$)$_6$O, we now shift our focus to the Cu-doped case.
According to experiment~\cite{Lee2023_2}, the LK-99 possesses a chemical formula
of Pb$_{10-x}$Cu$_x$(PO$_4$)$_6$O (0.9$<$x$<$1.1) with the Cu atoms substituting the Pb1 positions~\cite{Lee2023_2}.
For simplicity, here we considered $x$=1, which was obtained by substituting two Pb atoms with two Cu atoms in the $P\bar{3}$-Pb$_{10}$(PO$_4$)$_6$O superstructure.
We examined all possible configurations for Cu substituting the Pb sites.
Eventually, forty-four symmetry-inequivalent configurations were obtained (see Supplementary Material Fig.~S2~\cite{SM}).
The computed substitutional formation energies for all the configurations are displayed in Fig.~\ref{Fig3_Ef}.
Let us begin by examining the general trend revealed by these magnetically collinear DFT+$U$ (without ESB) data,
and then delve into a detailed discussion about the specific configurations.
First, one can see that the substitutional formation energies are overall large.
The lowest-energy configuration (Cfg1) with one Cu atom at a Pb2 site
and the other Cu atom at the neighboring Pb1 site is insulating (see Supplementary Material Fig.~S3~\cite{SM})
and exhibits a substitutional formation energy of 2.65 eV/Cu.
These results indicate the difficulty in incorporating Cu atoms at the Pb sites, in line with experimental implications~\cite{arXiv:2308.05776}.
In general, the doped Cu atoms exhibit a preference for substituting the Pb2 sites than the Pb1 sites,
which is manifested by much larger substitutional formation energies
for the configurations with two Cu atoms at the Pb1 sites than those with two Cu atoms at the Pb2 sites (see Fig.~\ref{Fig3_Ef} and  Supplementary Material Fig.~S2~\cite{SM}).
Additionally, an interesting observation is that the two Cu atoms
have a tendency to approach each other, indicating a preference for Cu-Cu interactions or clustering within the lattice.
Second, it can be seen that the energetically more favorable configurations are mostly insulating.
The abrupt increase appearing in DFT+$U$ calculated substitutional formation energies
corresponds to the boundary between insulators and metals.
Third, for the configurations where both two Cu atoms are at the Pb1 sites,
the configuration Cfg28 is found to exhibit the lowest substitutional formation energy.
In this structure, the two Cu atoms form a hexagonal lattice in the $ab$-plane
and are related to each other by an inversion symmetry, thereby preserving the $P\bar{3}$ symmetry of the system.
The magnetically collinear DFT+$U$ calculations without ESB predict Cfg28 to be metallic,
with four spin-polarized flat bands around the Fermi level,
similar to what was observed for the $P6_3/m$-Pb$_{9}$Cu(PO$_4$)$_6$O structure in previous DFT+$U$  calculations.
The in-plane hexagonal lattice results in time-reversal symmetry breaking induced Weyl points at K and H points [see Fig.~\ref{Fig4_all_bands}(b)].

It should be noted that the long-range ferromagnetic (FM) magnetic ordering was adopted for above DFT+$U$ calculations.
Despite being lower in energy as compared to the antiferromagnetic  and nonmagnetic (NM) configurations,
the FM ordered configuration without ESB may trap in a high-symmetry local energy minima.
To check this point, we performed electronic optimization
using the conjugate gradient algorithm~\cite{RevModPhys.64.1045} and fully accounted for the ESB.
Interestingly, with the ESB, the band gap is opened, leading to an insulating state [compare Figs.~\ref{Fig4_all_bands}(b) and (a)].
Meanwhile, the total energy is significantly lowered by 0.57 eV per unit cell.
In addition, we also conducted magnetically noncollinear DFT+$U$ calculations with the ESB,
yielding similar insulating band structure [compare Figs.~\ref{Fig4_all_bands}(c) and (b)]
and a lowered system energy (see Fig.~\ref{Fig3_Ef}).
The effect of spin-obit coupling (SOC) was also investigated,
and as expected, the SOC exhibits negligible effect [compare Figs.~\ref{Fig4_all_bands}(c) and (d)].
Furthermore, we also reexamined the electronic structure of $P6_3/m$-Pb$_{9}$Cu(PO$_4$)$_6$O.
We recall that the $P6_3/m$ structure
is the one that was studied by previous calculations in literature~\cite{JMST2023,arXiv:2307.16892,Liang1,arXiv:2308.00698,arXiv:2308.01135,arXiv:2308.01315,
arXiv:2308.02469,arXiv:2308.03218,OganovDMFT,Liang2,PhilippDMFT,arXiv:2308.05124,arXiv:2308.05143,arXiv:2308.05618,Liang3}.
For this structure, the Cu atom is placed at the energetically unfavorable Pb1 site.
Previous magnetically collinear DFT+$U$ calculations without the ESB predicted a metallic state with two flat bands around the Fermi level.
However, with the ESB, the band gap is also opened (see Supplementary Material Figs.~S4~\cite{SM}).
All these results suggest that the previously reported flat-band metallic state
is a result of an artifact of DFT+$U$ calculations without considering the ESB.
As a complement to our work, Swift and Lyons~\cite{arXiv:2308.08458} recently commented
that the incorrect prediction by DFT+$U$ calculations
originates from the overestimation of the energy of the O-$2p$ states in the valence band,
which can be remedied by hybrid functional calculations.
We note that similar situations apply to the configurations Cfg29, Cfg30 and Cfg31,
for which noncollinear DFT+$U$ calculations with the ESB also result in gap opening (see Fig.~\ref{Fig3_Ef}).
We also checked the $U$ dependence and found that the smaller value of $U$= 3.0 eV does not change the conclusion.

\begin{figure}
\begin{center}
\includegraphics[width=0.48\textwidth, clip]{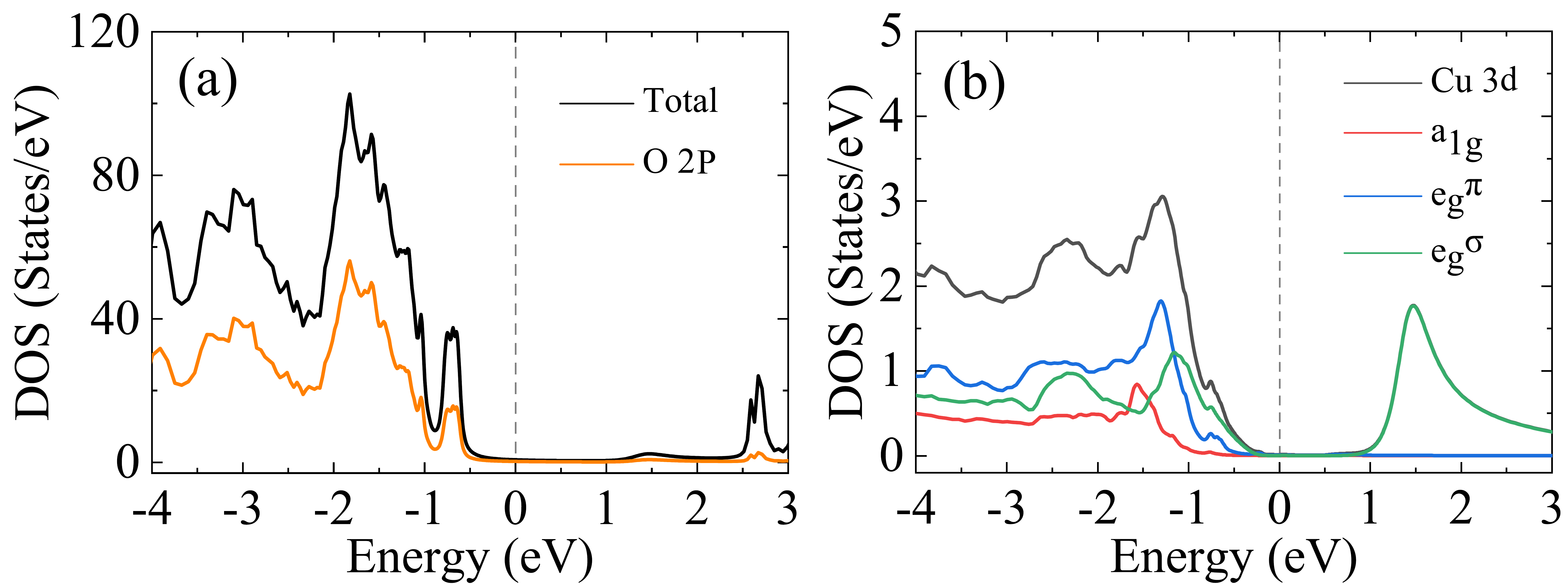}
\end{center}
\caption{DFT+DMFT calculated density of states for Cfg28-$P\bar{3}$-Pb$_{9}$Cu(PO$_4$)$_6$O at 387 K.
A paramagnetic setup is adopted.
}
\label{Fig5_DMFT}
\end{figure}

Before closing this section, we would like to briefly discuss the effect of dynamical electronic correlations.
As introduced in the Introduction section, DFT+DMFT calculations~\cite{OganovDMFT,PhilippDMFT}
showed that the stoichiometric Pb$_{9}$Cu(PO$_4$)$_6$O compound is a Mott insulator.
Again, the employed structure relies on the $P6_3/m$ structure with Cu atom at the Pb1 site.
Here, we conducted paramagnetic DFT+DMFT calculations at 387 K for the Cfg28-$P\bar{3}$-Pb$_{9}$Cu(PO$_4$)$_6$O structure.
The local interactions were treated at the density-density level,
with the Hund's coupling $J=$ 0.07$U$ and $U$= 7.0 eV.
The large $U$ value used here is due to the fact that
we employed a full-potential DMFT implementation~\cite{haule2010dynamical}
that incorporates the states spanning a large energy window around the Fermi level.
These interaction parameters are comparable to cuprates ($U$= 7.0 eV within the $d$-$p$ model~\cite{PhysRevB.91.125142})
and also to $P6_3/m$-Pb$_{9}$Cu(PO$_4$)$_6$O  ($U$= 5.67 eV within the $d$-$p$ model~\cite{PhilippDMFT}).
The DFT+DMFT calculated DOSs for Cfg28-$P\bar{3}$-Pb$_{9}$Cu(PO$_4$)$_6$O are displayed in Fig.~\ref{Fig5_DMFT}.
It can be seen that, in contrast to the prediction by the magnetically collinear DFT+$U$ without ESB,
the paramagnetic DFT+DMFT predicts an insulating state for the Cfg28-$P\bar{3}$-Pb$_{9}$Cu(PO$_4$)$_6$O compound,
similar to the $P6_3/m$-Pb$_{9}$Cu(PO$_4$)$_6$O compound~\cite{OganovDMFT,PhilippDMFT}.
All these results conclude that the stoichiometric Cfg28-$P\bar{3}$-Pb$_{9}$Cu(PO$_4$)$_6$O structure
is a robust insulator at both low and high temperatures.

\section{Conclusions}

In conclusion, we have revisited the electronic structure of Pb$_{9}$Cu(PO$_4$)$_6$O
based on the energetically more favorable P$\bar{3}$-Pb$_{10}$(PO$_4$)$_6$O structure,
fully taking into account the ESB.
We examine all possible configurations for Cu substituting the Pb sites
and demonstrate that the doped Cu atoms prefer to substitute the Pb2 sites rather than the Pb1 sites.
In both cases, the calculated substitutional formation energies are large,
posing great challenge for experiments to dope Cu atoms at the Pb sites.
It is found that the energetically favorable Cu-doped configurations are insulating,
in lines with many experimental observations
showing insulating transport behaviors in synthesized
samples~\cite{arXiv:2308.06589, arXiv:2307.16402, arXiv:2308.05778, wen_arXiv:2303.08759, arXiv:2308.06256}.
Our calculations show that the conducting state can only be obtained in a few Cu-doped configurations
with Cu at the Pb1 sites. However, due to their larger substitutional formation energies,
it would be difficult to obtain a conducting state for stoichiometric Pb$_{9}$Cu(PO$_4$)$_6$O
just by doping Cu atoms from a thermodynamical perspective.
Importantly, we find that the magnetically collinear DFT+$U$ method without the electronic symmetry breaking
predicts an incorrect flat-band metallic state for some Cu-doped configurations with Cu atoms at the Pb1 sites.
These also include the $P6_3/m$-Pb$_{9}$Cu(PO$_4$)$_6$O structure that previous DFT+$U$ calculations rely on.
By accounting for the ESB,
the correct insulating electronic state is obtained.
This has been demonstrated by simply incorporating the ESB in magnetically collinear DFT+$U$ calculations,
or through magnetically noncollinear DFT+$U$ calculations and DFT+DMFT calculations.
Our work emphasizes the importance of symmetry breaking in electronic structure calculations
and reconciles the inconsistency between previous DFT+$U$ and DFT+DMFT calculations.

\emph{Note added}. During the preparation of this manuscript, we became aware of a similar work
that also emphasizes the importance of symmetry breaking~\cite{arXiv:2308.07295}.

\section{Acknowledgments}
The authors thank Dr. Krivovichev for sharing the structural data of $P\bar{3}$-Pb$_{10}$(PO$_4$)$_6$O.

\bibliographystyle{apsrev4-1}
\bibliography{Reference} 

\end{document}


\title{Supplementary Material to \\
		``Symmetry breaking induced insulating electronic state in Pb$_{9}$Cu(PO$_4$)$_6$O"}

\author{Jiaxi Liu}
\thanks{These authors contribute equally to this work.}
\affiliation{%
Shenyang National Laboratory for Materials Science, Institute of Metal Research, Chinese Academy of Sciences, 110016 Shenyang, China.
}%
\affiliation{%
School of Materials Science and Engineering, University of Science and Technology of China, 110016 Shenyang, China.
}%

\author{Tianye Yu}
\thanks{These authors contribute equally to this work.}
\affiliation{%
Shenyang National Laboratory for Materials Science, Institute of Metal Research, Chinese Academy of Sciences, 110016 Shenyang, China.
}%

\author{Jiangxu Li}
\email{jxli15s@imr.ac.cn}
\affiliation{%
Shenyang National Laboratory for Materials Science, Institute of Metal Research, Chinese Academy of Sciences, 110016 Shenyang, China.
}%

\author{Jiantao Wang}
\affiliation{%
Shenyang National Laboratory for Materials Science, Institute of Metal Research, Chinese Academy of Sciences, 110016 Shenyang, China.
}%
\affiliation{%
School of Materials Science and Engineering, University of Science and Technology of China, 110016 Shenyang, China.
}%

\author{Junwen Lai}
\affiliation{%
Shenyang National Laboratory for Materials Science, Institute of Metal Research, Chinese Academy of Sciences, 110016 Shenyang, China.
}%
\affiliation{%
School of Materials Science and Engineering, University of Science and Technology of China, 110016 Shenyang, China.
}%

\author{Yan Sun}
\affiliation{%
Shenyang National Laboratory for Materials Science, Institute of Metal Research, Chinese Academy of Sciences, 110016 Shenyang, China.
}%

\author{Xing-Qiu Chen}%
\affiliation{%
Shenyang National Laboratory for Materials Science, Institute of Metal Research, Chinese Academy of Sciences, 110016 Shenyang, China.
}%

\author{Peitao Liu}%
\email{ptliu@imr.ac.cn}
\affiliation{%
Shenyang National Laboratory for Materials Science, Institute of Metal Research, Chinese Academy of Sciences, 110016 Shenyang, China.
}%

\maketitle

	\begin{figure}
		\begin{center}
			\includegraphics[width=0.95\textwidth, clip]{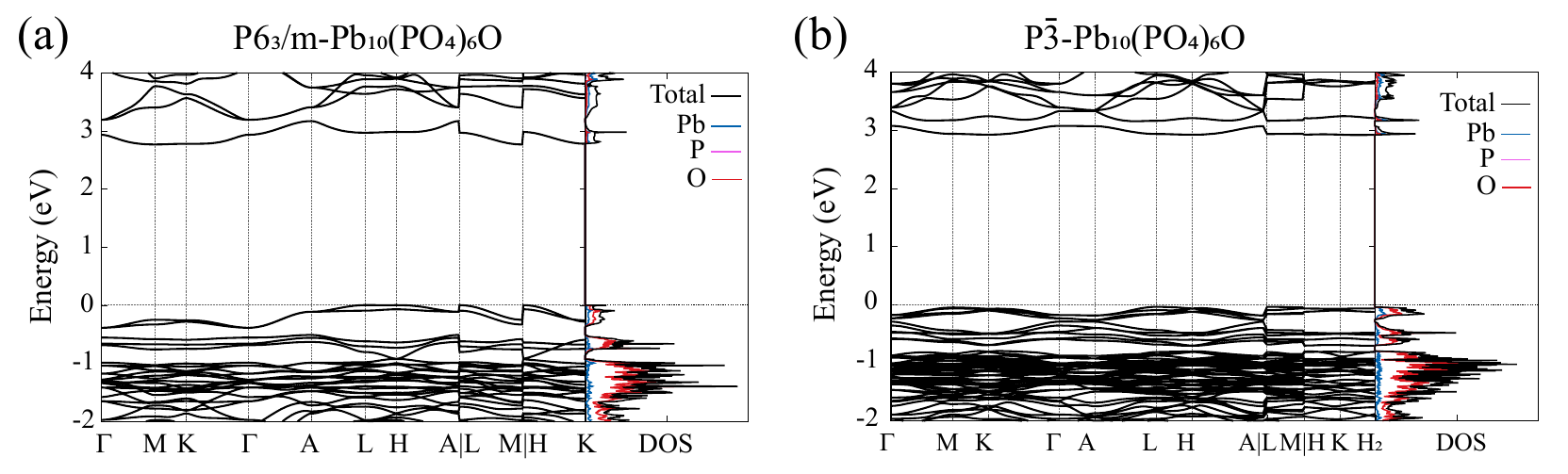}
		\end{center}
		\caption{Calculated electronic band structure and density of states (DOSs) of (a) $P6_3/m$-Pb$_{10}$(PO$_4$)$_6$O and (b) $P\bar{3}$-Pb$_{10}$(PO$_4$)$_6$O.
		}
		\label{FigS1}
	\end{figure}

	\begin{figure}
		\begin{center}
			\includegraphics[width=0.65\textwidth, clip]{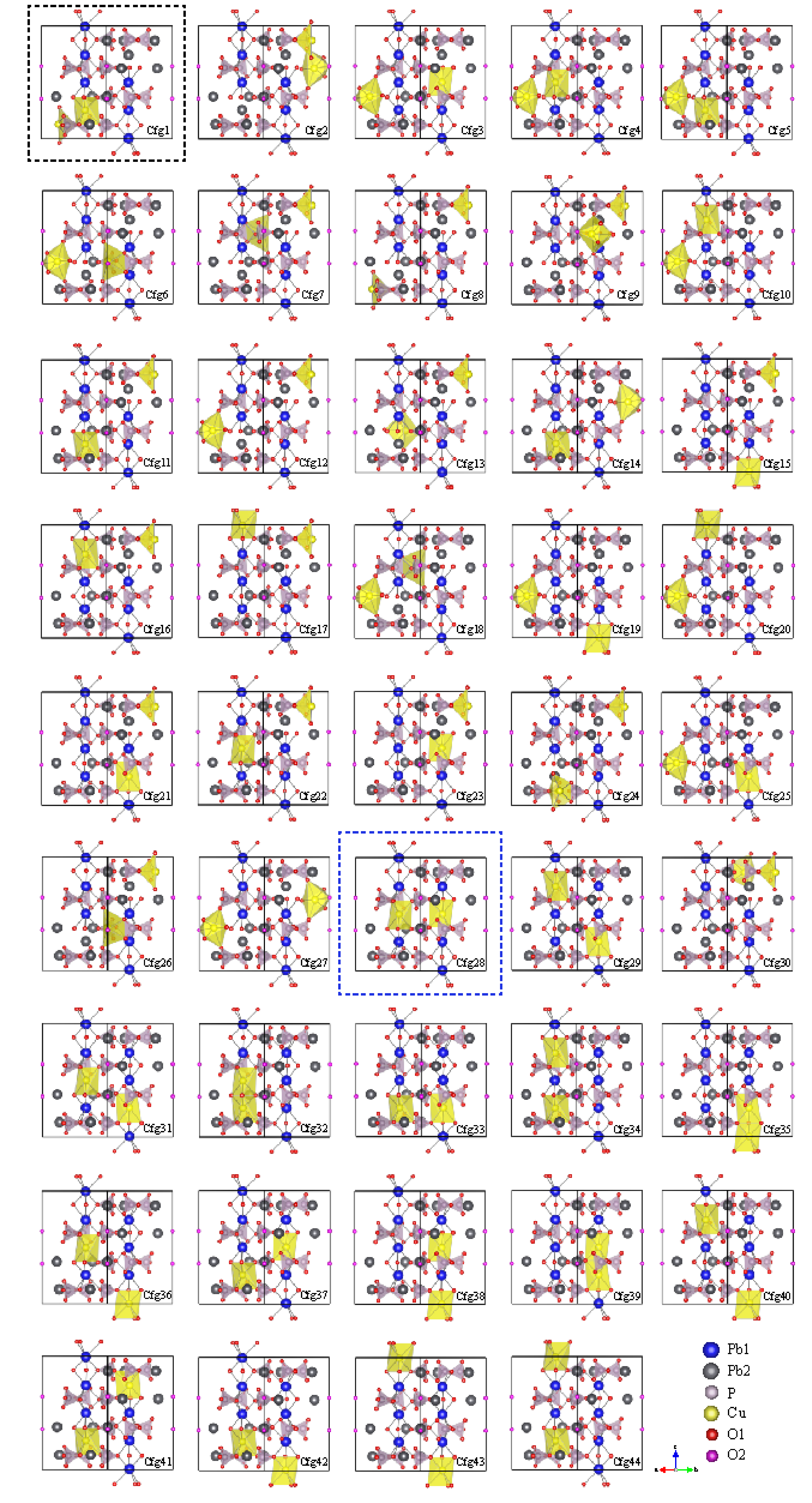}
		\end{center}
		\caption{Different configurations of $P\bar{3}$-Pb$_{9}$Cu(PO$_4$)$_6$O with two Cu atoms substituting two Pb sites.
The configurations are arranged in ascending magnetically collinear DFT+$U$ (without electronic symmetry breaking) calculated total energies.
The Cu atoms are marked in yellow polyhedra.
The lowest-energy configuration (Cfg1) is highlighted in a black dashed rectangle.
The lowest-energy configuration where both two Cu atoms are at the Pb1 sites (Cfg28)  is highlighted in a blue dashed rectangle.
		}
		\label{FigS2}
	\end{figure}

\begin{table*}
\caption{Computed Cu$\rightarrow$Pb substitutional formation energies ($E_f$, in eV)
and band gap (in eV) for forty-four symmetry-inequivalent configurations
by magnetically collinear DFT+$U$ (without electronic symmetry breaking) and DFT+$U$+SOC calculations.
$M$(Cu1) and  $M$(Cu1) (in $\mu_\text{B}$)  indicate local magnetic moments of two Cu atoms predicted by collinear DFT+$U$ calculations.
$M_x$, $M_y$, and $M_z$ (in $\mu_\text{B}$) denote the total magnetic moment of unit cell
along the $x$, $y$, and $z$ axis, respectively, predicted by noncollinear DFT+$U$ calculations.
}
\begin{ruledtabular}
\begin{tabular}{rcccccccccc}
& \multicolumn{4}{c}{Collinear DFT+$U$ (without ESB)} & & \multicolumn{5}{c}{DFT+$U$+SOC}  \\
\hline
 Cfg & $E_f$  & $E_g$  & $M$(Cu1)  & $M$(Cu2) & &  $E_f$ & $E_g$  & $M_x$  & $M_y$ & $M_z$  \\
 \hline
 1 &   2.65 &    0.70  &  0.67 &  0.70  & & 2.65  &  0.74  &  1.98 &  -0.06 & -0.03  \\
 2 &   2.65 &    0.72  & -0.65 &  0.63  & & 2.66  &  0.77  & -1.97 &   0.01 &  0.09  \\
 3 &   2.66 &    0.83  & -0.66 & -0.73  & & 2.65  &  0.86  & -1.99 &  -0.00 &  0.01  \\
 4 &   2.66 &    1.15  &  0.66 & -0.71  & & 2.66  &  1.20  &  1.99 &  -0.01 & -0.00  \\
 5 &   2.74 &    1.09  &  0.68 & -0.70  & & 2.74  &  1.12  &  1.99 &   0.00 & -0.02  \\
 6 &   2.74 &    1.25  & -0.68 &  0.59  & & 2.74  &  1.22  &  1.99 &  -0.01 & -0.01  \\
 7 &   2.74 &    0.70  &  0.64 & -0.66  & & 2.74  &  0.76  &  1.98 &  -0.06 & -0.04  \\
 8 &   2.77 &    0.65  &  0.64 & -0.64  & & 2.78  &  0.70  &  1.98 &  -0.09 & -0.06  \\
 9 &   2.77 &    0.69  &  0.63 &  0.65  & & 2.77  &  0.76  &  1.98 &  -0.03 & -0.06  \\
10 &   2.77 &    1.19  &  0.67 & -0.71  & & 2.76  &  1.23  &  1.99 &   0.04 & -0.05  \\
11 &   2.79 &    0.57  &  0.64 &  0.70  & & 2.78  &  0.63  &  1.98 &  -0.05 & -0.03  \\
12 &   2.80 &    0.54  &  0.64 & -0.63  & & 2.80  &  0.60  &  1.98 &  -0.07 &  0.06  \\
13 &   2.80 &    0.64  &  0.65 &  0.65  & & 2.80  &  0.73  &  1.98 &  -0.03 & -0.01  \\
14 &   2.81 &    0.96  &  0.66 & -0.70  & & 2.81  &  1.02  &  1.99 &   0.02 &  0.02  \\
15 &   2.82 &    1.05  &  0.65 & -0.70  & & 2.80  &  1.00  &  1.99 &  -0.03 & -0.02  \\
16 &   2.83 &    0.52  &  0.66 &  0.72  & & 2.82  &  0.57  & -1.98 &  -0.05 & -0.09  \\
17 &   2.83 &    0.64  &  0.64 & -0.71  & & 2.82  &  0.71  &  1.99 &  -0.02 & -0.03  \\
18 &   2.84 &    1.15  &  0.64 &  0.64  & & 2.86  &  1.16  &  1.98 &  -0.08 &  0.07  \\
19 &   2.85 &    0.88  &  0.67 & -0.70  & & 2.84  &  0.93  &  1.99 &   0.02 &  0.02  \\
20 &   2.85 &    0.98  &  0.67 &  0.72  & & 2.84  &  1.02  &  1.98 &   0.01 &  0.02  \\
21 &   2.86 &    0.55  &  0.64 &  0.71  & & 2.85  &  0.62  &  1.99 &  -0.06 & -0.03  \\
22 &   2.86 &    0.42  &  0.64 &  0.71  & & 2.85  &  0.48  &  1.98 &  -0.01 & -0.00  \\
23 &   2.87 &    0.45  & -0.64 & -0.72  & & 2.85  &  0.50  & -1.98 &  -0.02 &  0.00  \\
24 &   2.90 &    0.61  &  0.64 & -0.71  & & 2.91  &  0.67  &  1.99 &  -0.04 & -0.03  \\
25 &   2.91 &    1.20  &  0.65 &  0.70  & & 2.90  &  1.25  &  1.99 &   0.01 & -0.01  \\
26 &   2.92 &    0.53  & -0.64 &  0.64  & & 2.94  &  0.58  & -1.87 &  -0.17 & -0.62  \\
27 &   2.93 &    1.24  &  0.64 & -0.64  & & 2.96  &  1.19  &  1.98 &  -0.01 & -0.05  \\
28 &   3.27 &    Metal & -0.58 & -0.58  & & 2.88  &  0.62  & -0.05 &   0.01 & -1.97  \\
29 &   3.33 &    Metal & -0.52 & -0.52  & & 2.98  &  0.48  &  0.67 &   0.01 &  1.86  \\
30 &   3.35 &    Metal &  0.01 & -0.00  & & 2.87  &  0.60  &  1.98 &  -0.01 & -0.05  \\
31 &   3.41 &    Metal & -0.50 & -0.52  & & 3.12  &  0.25  &  0.55 &  -0.06 & -1.89  \\
32 &   3.47 &    Metal & -0.42 & -0.35  & & 3.30  &  Metal &  1.83 &  -0.05 &  0.12  \\
33 &   3.48 &    Metal & -0.12 & -0.03  & & 3.30  &  Metal &  1.40 &   0.07 &  0.58  \\
34 &   3.48 &    Metal & -0.65 & -0.41  & & 3.21  &  Metal &  1.54 &   0.01 & -0.77  \\
35 &   3.49 &    Metal & -0.28 & -0.44  & & 3.37  &  Metal &  1.68 &   0.12 & -0.64  \\
36 &   3.49 &    Metal &  0.38 &  0.58  & & 3.20  &  Metal &  1.84 &   0.05 &  0.57  \\
37 &   3.53 &    Metal & -0.54 & -0.41  & & 3.29  &  Metal &  1.54 &   0.01 & -0.74  \\
38 &   3.55 &    Metal &  0.40 &  0.63  & & 3.25  &  Metal &  1.89 &   0.03 &  0.38  \\
39 &   3.57 &     0.08 & -0.61 & -0.47  & & 3.28  &  0.38  &  0.42 &   0.05 & -1.93  \\
40 &   3.60 &    Metal &  0.24 &  0.62  & & 3.45  &  Metal &  1.71 &   0.15 & -0.08  \\
41 &   3.62 &    Metal & -0.48 & -0.48  & & 3.30  &  Metal &  0.82 &   0.08 &  1.78  \\
42 &   3.73 &    Metal & -0.50 & -0.32  & & 3.52  &  Metal & -1.10 &  -0.40 &  1.35  \\
43 &   3.79 &    Metal & -0.33 & -0.33  & & 3.67  &  Metal &  1.54 &   0.11 &  0.04  \\
44 &   3.84 &    Metal &  0.34 &  0.51  & & 3.63  &  Metal &  1.81 &  -0.04 & -0.31  \\
\end{tabular}
\end{ruledtabular}
\label{TableS1}
\end{table*}

	\begin{figure}
		\begin{center}
			\includegraphics[width=0.80\textwidth, clip]{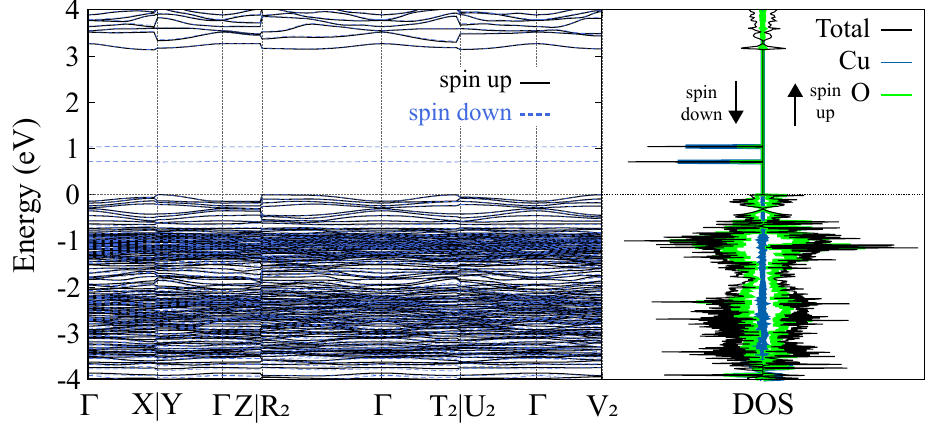}
		\end{center}
		\caption{Predicted electronic band structure of Cfg1-$P\bar{3}$-Pb$_{9}$Cu(PO$_4$)$_6$O
by magnetically collinear DFT+$U$ method with a ferromagnetic setup.
		}
		\label{FigS3}
	\end{figure}

	\begin{figure}
		\begin{center}
			\includegraphics[width=0.90\textwidth, clip]{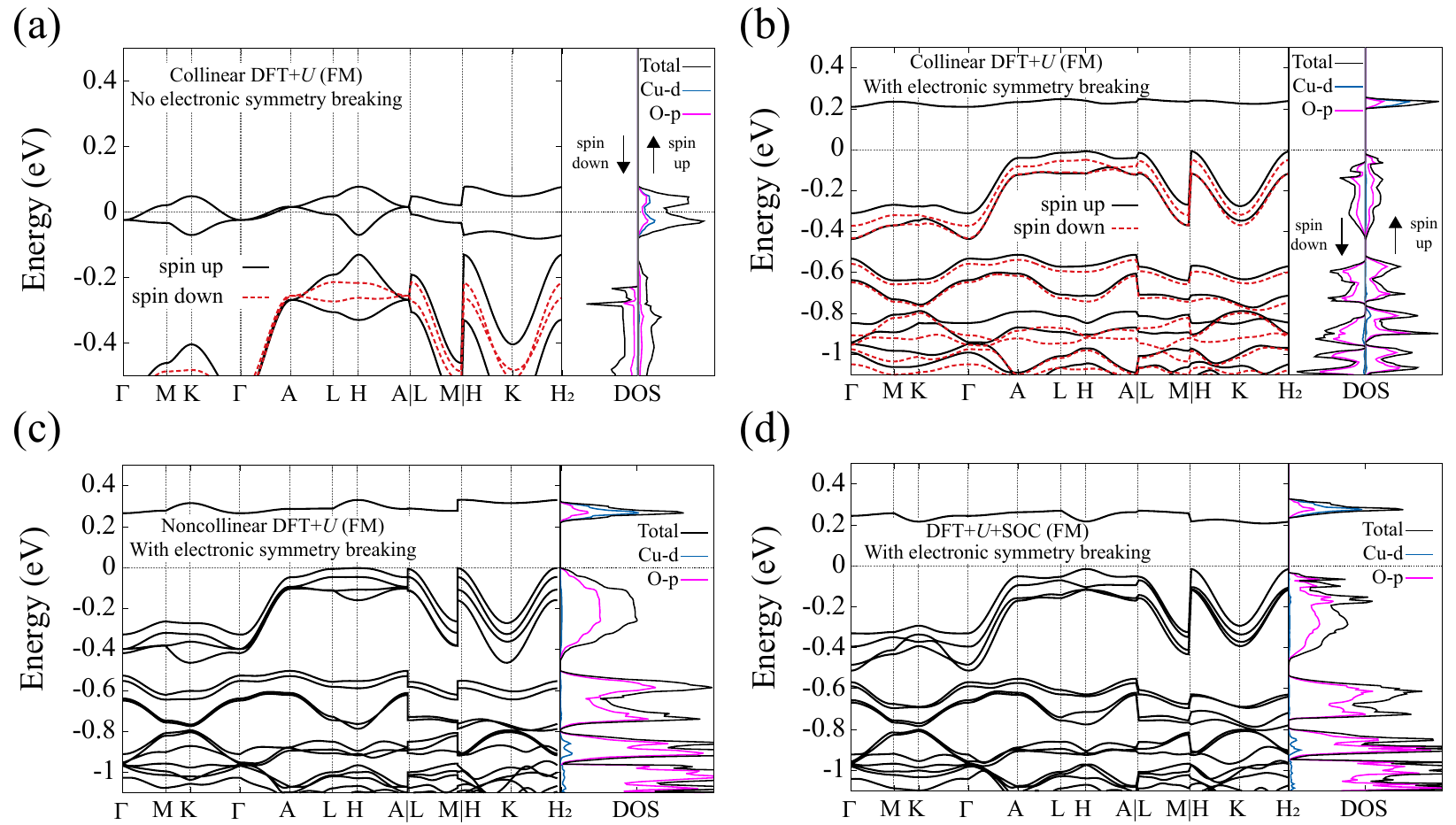}
		\end{center}
		\caption{Predicted electronic band structure and density of states for $P6_3/m$-Pb$_{9}$Cu(PO$_4$)$_6$O with Cu substituting the Pb1 site
by (a) magnetically collinear DFT+$U$ without electronic symmetry breaking (ESB),
(b) magnetically collinear DFT+$U$ with ESB,
(c) magnetically noncollinear DFT+$U$ with ESB,
and (d) DFT+$U$+SOC with ESB. A ferromagnetic setup is adopted for all the calculations.
Note that this structure is the one that was studied for previous calculations in literature.
One can see that with the ESB,  $P6_3/m$-Pb$_{9}$Cu(PO$_4$)$_6$O becomes insulating,
and accordingly, the system energy is lowered by 0.23 eV per unit cell.
		}
		\label{FigS4}
	\end{figure}

\clearpage

In the following, the structural data (in the VASP POSCAR format) for
$P6_3/m$-Pb$_{10}$(PO$_4$)$_6$O,
$P\bar{3}$-Pb$_{10}$(PO$_4$)$_6$O,
$P6_3/m$-Pb$_{9}$Cu(PO$_4$)$_6$O,
and $P\bar{3}$-Pb$_{9}$Cu(PO$_4$)$_6$O for configurations Cfg1 and Cfg28 are provided.

\begin{verbatim}
P63/m-Pb10(PO4)6O
   1.00000000000000
    10.0272251983801937    0.0000004282969099    0.0000000000000000
    -5.0136129701585634    8.6838315371159585    0.0000000000000000
     0.0000000000000000    0.0000000000000000    7.4884439295252694
   Pb   P    O
    10     6    25
Direct
  0.9961954751515520  0.7711081176220773  0.2466023428248363
  0.0051256466906082  0.2646237955944386  0.7469308423330361
  0.2288918823779227  0.2250873575294747  0.2466023428248363
  0.7353762044055614  0.7405018520961733  0.7469308423330361
  0.7749126574705301  0.0038045388484491  0.2466023428248363
  0.2594981629038315  0.9948743673093929  0.7469308423330361
  0.6666666870000029  0.3333333429999996  0.0019946548877812
  0.3333333129999971  0.6666666269999979  0.0070835047523587
  0.3333333129999971  0.6666666269999979  0.4870177106723759
  0.6666666870000029  0.3333333429999996  0.4926955212948201
  0.6237426224552181  0.5945665022417259  0.2473144588411174
  0.3668247798702637  0.3950350885859919  0.7473119078111026
  0.4054334977582741  0.0291760902134897  0.2473144588411174
  0.6049649114140081  0.9717897212842743  0.7473119078111026
  0.9708239397865128  0.3762574075447844  0.2473144588411174
  0.0282103087157282  0.6331752501297387  0.7473119078111026
  0.5004265938327563  0.6454684767698140  0.2471544138136679
  0.4690551250992812  0.3198857735149758  0.7473709905233434
  0.3545315232301860  0.8549581170629423  0.2471544138136679
  0.6801142264850242  0.1491693505843088  0.7473709905233434
  0.1450418529370552  0.4995733771672377  0.2471544138136679
  0.8508306194156958  0.5309448459007129  0.7473709905233434
  0.7296726113170990  0.6566362271067376  0.0789912972928590
  0.2597284814647836  0.3499464570056929  0.9142832095743145
  0.3433637728932624  0.0730363542103589  0.0789912972928590
  0.6500535429943071  0.9097820534590824  0.9142832095743145
  0.9269636747896470  0.2703274186829034  0.0789912972928590
  0.0902179755409236  0.7402715485352189  0.9142832095743145
  0.2594803082543393  0.3495916752009336  0.5805657087045191
  0.7297590515043311  0.6570435063739595  0.4155139080278474
  0.6504083247990664  0.9098886620534046  0.5805657087045191
  0.3429564936260405  0.0727155151303762  0.4155139080278474
  0.0901113669466014  0.7405197217456632  0.5805657087045191
  0.9272845138696297  0.2702409784956714  0.4155139080278474
  0.5445669592313251  0.4146327383607797  0.2475566182991216
  0.4719543955781518  0.5766699612335842  0.7470816551885875
  0.5853672316392178  0.1299342208705454  0.2475566182991216
  0.4233300087664063  0.8952843743445627  0.7470816551885875
  0.8700657491294521  0.4554330407686749  0.2475566182991216
  0.1047155956554349  0.5280456044218482  0.7470816551885875
  0.0000000000000000  0.0000000000000000  0.2453765386895981
\end{verbatim}

\begin{verbatim}
P-3-Pb10(PO4)6O
   1.00000000000000
    10.0186438101416524    0.0000004879006388    0.0000000000000000
    -5.0093223275519589    8.6763998071436266    0.0000000000000000
     0.0000000000000000    0.0000000000000000   15.0335087536342069
   Pb   P    O
    20    12    50
Direct
  0.2604505818353786  0.9980904878151762  0.8701625893356564
  0.7395494181646214  0.0019094871848182  0.1298374256643342
  0.0019094871848182  0.2623600640202000  0.8701625893356564
  0.9980904878151762  0.7376399059798047  0.1298374256643342
  0.7376399059798047  0.7395494181646214  0.8701625893356564
  0.2623600640202000  0.2604505818353786  0.1298374256643342
  0.9990464505833359  0.2251237170448519  0.3750573167479061
  0.0009535264166658  0.7748762979551458  0.6249427132520893
  0.7748762979551458  0.7739227475384709  0.3750573167479061
  0.2251237170448519  0.2260772374615243  0.6249427132520893
  0.2260772374615243  0.0009535264166658  0.3750573167479061
  0.7739227475384709  0.9990464505833359  0.6249427132520893
  0.3333333429999996  0.6666666870000029  0.0048460320591417
  0.6666666269999979  0.3333333129999971  0.9951539919408603
  0.3333333429999996  0.6666666870000029  0.2568487137004212
  0.6666666269999979  0.3333333129999971  0.7431512862995717
  0.3333333429999996  0.6666666870000029  0.5005750294769271
  0.6666666269999979  0.3333333129999971  0.4994249705230658
  0.3333333429999996  0.6666666870000029  0.7434956661587933
  0.6666666269999979  0.3333333129999971  0.2565043628412056
  0.3706368717444946  0.4009850766235630  0.3710604551486298
  0.6293631282555054  0.5990149533764324  0.6289395738513690
  0.5990149533764324  0.9696517651209220  0.3710604551486298
  0.4009850766235630  0.0303482048790755  0.6289395738513690
  0.0303482048790755  0.6293631282555054  0.3710604551486298
  0.9696517651209220  0.3706368717444946  0.6289395738513690
  0.0258536843246233  0.6277897473237815  0.8791416579380282
  0.9741463136753765  0.3722102526762185  0.1208583270619599
  0.3722102526762185  0.3980639390008420  0.8791416579380282
  0.6277897473237815  0.6019360609991509  0.1208583270619599
  0.6019360609991509  0.9741463136753765  0.8791416579380282
  0.3980639390008420  0.0258536843246233  0.1208583270619599
  0.8480079145157546  0.5219142353925221  0.8781021076039863
  0.1519920704842548  0.4780857346074825  0.1218978623960112
  0.4780857346074825  0.3260936791232325  0.8781021076039863
  0.5219142353925221  0.6739063208767675  0.1218978623960112
  0.6739063208767675  0.1519920704842548  0.8781021076039863
  0.3260936791232325  0.8480079145157546  0.1218978623960112
  0.4912519837651601  0.3473271043453892  0.3685536692684153
  0.5087480162348399  0.6526729246546168  0.6314463307315847
  0.6526729246546168  0.1439248784197673  0.3685536692684153
  0.3473271043453892  0.8560751505802315  0.6314463307315847
  0.8560751505802315  0.5087480162348399  0.3685536692684153
  0.1439248784197673  0.4912519837651601  0.6314463307315847
  0.0820302768369743  0.7423205923301310  0.7991833875554022
  0.9179697461630241  0.2576794076698619  0.2008165974445859
  0.2576794076698619  0.3397096915068403  0.7991833875554022
  0.7423205923301310  0.6602903384931622  0.2008165974445859
  0.6602903384931622  0.9179697461630241  0.7991833875554022
  0.3397096915068403  0.0820302768369743  0.2008165974445859
  0.2740660668670500  0.3597436273398955  0.2826757982949886
  0.7259339331329429  0.6402563726601045  0.7173242017050114
  0.6402563726601045  0.9143224395271474  0.2826757982949886
  0.3597436273398955  0.0856775604728597  0.7173242017050114
  0.0856775604728597  0.7259339331329429  0.2826757982949886
  0.9143224395271474  0.2740660668670500  0.7173242017050114
  0.2538179418529722  0.3200039504844838  0.4484565062707304
  0.7461820581470278  0.6799960495155162  0.5515434637292813
  0.6799960495155162  0.9338139923684921  0.4484565062707304
  0.3200039504844838  0.0661860076315079  0.5515434637292813
  0.0661860076315079  0.7461820581470278  0.4484565062707304
  0.9338139923684921  0.2538179418529722  0.5515434637292813
  0.0863980332698944  0.7252466484728330  0.9657131684522895
  0.9136019737301098  0.2747533515271670  0.0342868385477075
  0.2747533515271670  0.3611513777970643  0.9657131684522895
  0.7252466484728330  0.6388486222029357  0.0342868385477075
  0.6388486222029357  0.9136019737301098  0.9657131684522895
  0.3611513777970643  0.0863980332698944  0.0342868385477075
  0.1055713392410169  0.5260830233551488  0.8717859358306299
  0.8944286457589996  0.4739169476448453  0.1282140491693582
  0.4739169476448453  0.5794882718858787  0.8717859358306299
  0.5260830233551488  0.4205117281141284  0.1282140491693582
  0.4205117281141284  0.8944286457589925  0.8717859358306299
  0.5794882718858787  0.1055713392410169  0.1282140491693582
  0.4516102949690648  0.5800140937606102  0.3825357045253526
  0.5483896750309256  0.4199859062393827  0.6174642664746415
  0.4199859062393827  0.8715962312084500  0.3825357045253526
  0.5800140937606102  0.1284037987915525  0.6174642664746415
  0.1284037987915525  0.5483896750309256  0.3825357045253526
  0.8715962312084500  0.4516102949690648  0.6174642664746415
  0.0000000000000000  0.0000000000000000  0.3561542961106241
  0.0000000000000000  0.0000000000000000  0.6438457338893784
\end{verbatim}

\begin{verbatim}
P63/m-Pb9Cu(PO4)6O
	1.00000000000000
	9.9306823821051555   -0.0000014818366361    0.0000000000000000
	-4.9653399077953200    8.6002239607356703   -0.0000000000000000
	0.0000000000000000    0.0000000000000000    7.4109945172796605
	Pb   Cu   P    O
	9    1    6    25
	Direct
	0.9986807449642078  0.7699202447216552  0.2474756380877509
	0.9977014382162520  0.2577003216124319  0.7545056960306785
	0.2300797552783452  0.2287605002425670  0.2474756380877509
	0.7422996783875685  0.7400011176038168  0.7545056960306785
	0.7712395147574378  0.0013192690357936  0.2474756380877509
	0.2599988973961879  0.0022985757837491  0.7545056960306785
	0.3333333129999971  0.6666666269999979  0.0102252768912011
	0.3333333129999971  0.6666666269999979  0.4962563424132084
	0.6666666870000029  0.3333333429999996  0.5217140548665616
	0.6666666870000029  0.3333333429999996  0.0634955699714910	
	0.6238378061607577  0.5940989414527301  0.2330614375419467
	0.3715115645822017  0.3913852574380640  0.7499251076346213
	0.4059010585472700  0.0297388347080183  0.2330614375419467
	0.6086147425619355  0.9801263371441329  0.7499251076346213
	0.9702611952919840  0.3761622238392447  0.2330614375419467
	0.0198736928558696  0.6284884654178010  0.7499251076346213
	0.4970341950086190  0.6419127807282180  0.2504819573050943
	0.4722574174374839  0.3124116086150668  0.7489819492636393
	0.3580872192717823  0.8551214142804011  0.2504819573050943
	0.6875883913849331  0.1598458078224275  0.7489819492636393
	0.1448785557195963  0.5029657759913753  0.2504819573050943
	0.8401541621775773  0.5277425535625101  0.7489819492636393
	0.7465890784344098  0.6974373665162791  0.0877326196589775
	0.2518710331486960  0.3302579489920529  0.9082269759127742
	0.3025626334837207  0.0491516819181211  0.0877326196589775
	0.6697420510079468  0.9216131131566275  0.9082269759127742
	0.9508483470818850  0.2534109515655927  0.0877326196589775
	0.0783869158433784  0.7481289968513065  0.9082269759127742
	0.2751067320564097  0.3631086026138700  0.5737407252403873
	0.7132743878686607  0.6201447197159712  0.4140528338576862
	0.6368913973861298  0.9119981584425314  0.5737407252403873
	0.3798552802840290  0.0931296381526946  0.4140528338576862
	0.0880018705574747  0.7248932979435926  0.5737407252403873
	0.9068703908473114  0.2867256421313418  0.4140528338576862
	0.5441603686812513  0.4191736441937305  0.1764358135764958
	0.4780900223868556  0.5733419996779769  0.7722935441387523
	0.5808263258062671  0.1249867244875282  0.1764358135764958
	0.4266579703220141  0.9047479627088743  0.7722935441387523
	0.8750132455124694  0.4558396313187486  0.1764358135764958
	0.0952520072911234  0.5219099776131445  0.7722935441387523
	-0.0000000000000000  0.0000000000000000  0.2917658511111268
\end{verbatim}

\begin{verbatim}
Cfg1-P-3-Pb9Cu(PO4)6O
   1.00000000000000
   1.00000000000000
     9.9235770926455658   -0.0628424665429319    0.0546293364813921
    -5.0166686106767067    8.7957546553353918   -0.0622639420571632
     0.0839886958645425   -0.0603624921551655   15.0090921678528773
   Pb   Cu   P    O
    18     2    12    50
Direct
  0.2597257225536183  0.9948707464628939  0.8677507343137520
  0.3307904393185126  0.6593637076880157  0.7442331967796108
  0.9982224924180798  0.2678375703020492  0.8723773364283929
  0.0097765102872813  0.7404239464715090  0.1286674886274071
  0.7377375693926851  0.7350962364825904  0.8743201702569365
  0.2818382265036163  0.2698697127731933  0.1303135832095705
  0.0126937694316851  0.2323445581317003  0.3862248177784835
  0.9987448931316791  0.7686042651948100  0.6298416336816004
  0.7826560211397222  0.7694710274504430  0.3866628559073106
  0.2161241509257721  0.2125040894921284  0.6322957352615219
  0.2279652078464949  0.0070395536694789  0.3815997092683148
  0.7713258940191778  0.9938998127001071  0.6343199401254083
  0.3516604375239467  0.6799069068323718  0.0044468314534498
  0.6638415079981783  0.3230868332152070  0.9965799630905465
  0.3286956603412747  0.6636824410716997  0.2531336451643540
  0.6623514780495441  0.3307090423995689  0.7378980161318438
  0.3444685931793288  0.6694929376698582  0.5016018493303278
  0.6614742096575696  0.3286933785033028  0.4871938202104076
  0.7160803740177570  0.0835702425274363  0.2127031724105848
  0.7009611361179395  0.4397118367030828  0.2499753454361198
  0.3931571713754778  0.4084714899845991  0.3567876086630548
  0.6295145548883596  0.5921290004880291  0.6275997585492519
  0.5937297722270287  0.9636073113157053  0.3841426628744671
  0.3985172105713559  0.0235316617050287  0.6236193782087938
  0.0234840787911210  0.6269624323762244  0.3626823758212510
  0.9744815432651635  0.3630827808717640  0.6227540989094962
  0.0227437826665806  0.6264154760475904  0.8756494764771645
  0.9596578917537784  0.3836029234889864  0.1244745969942969
  0.3674746143082999  0.3913136919429974  0.8730266230233568
  0.6315063056206682  0.6017705224206225  0.1224118891390873
  0.6004474532460904  0.9668231238965035  0.8760578819496985
  0.3996949305500479  0.0260500921325075  0.1228248560803564
  0.8436669690840475  0.5198669642651694  0.8738595513339575
  0.1353046324592313  0.4996598774625127  0.1313312841375236
  0.4722097307875899  0.3182096870453321  0.8683641545054712
  0.4936240451494314  0.6319849298650624  0.1221101382045404
  0.6739467780725832  0.1427585709949497  0.8807595169663003
  0.3367301713752582  0.8509884944963630  0.1206460891415801
  0.5281933119320144  0.3744563585960279  0.3383617974635342
  0.5143393702837074  0.6530047188207107  0.6319276993002489
  0.6734829920728345  0.1390328089032593  0.3845912038664139
  0.3518895027666531  0.8532205888744002  0.6293541948371910
  0.8481379875354733  0.5011139453156161  0.3497042350069322
  0.1472102407382607  0.4900255784227880  0.6297960745945090
  0.0796586166658102  0.7417315134776175  0.7964390591582742
  0.9155401489587618  0.2464629576471893  0.1878400912184901
  0.2522407288155790  0.3391023952863392  0.7926167458089708
  0.7554062791628553  0.6921125436085163  0.1949948123208500
  0.6651986758992621  0.9184678326712614  0.7943871668954614
  0.3049940399205795  0.0617504194249108  0.1933178370955773
  0.2847960620990904  0.3722866692970612  0.2736292805206944
  0.7311960047690960  0.6334554543299191  0.7142671490781467
  0.6300285780927766  0.9154348306479960  0.2888397909219051
  0.3618688819138995  0.0825626732392450  0.7134853428543835
  0.0923643354161143  0.7231367741055266  0.2759666029755081
  0.9121989586178856  0.2709638966591683  0.7118851021796146
  0.2892465152602881  0.3054503668616704  0.4337926533079965
  0.7458006222762492  0.6681192181577060  0.5483939971540153
  0.6583934135041005  0.9018720948736245  0.4564296582769884
  0.3018974973168511  0.0465861525503897  0.5501841667336365
  0.0357986669652348  0.7395178576251382  0.4377545018182971
  0.9589745925388300  0.2432919038483163  0.5502435382761490
  0.0817869640894529  0.7235689856845369  0.9625270768989154
  0.9121169330519052  0.3271578815274694  0.0277680632172235
  0.2676891908380696  0.3501005042424765  0.9596090489143450
  0.7111138614605892  0.6231254606386329  0.0305522686334143
  0.6275727186074320  0.8939898442951559  0.9595819017349569
  0.3806859809277370  0.0878909939514330  0.0318210643906482
  0.1056044761633643  0.5284591642788072  0.8653751427840817
  0.8636596060526998  0.4650778246815293  0.1543227434908232
  0.4712703997636609  0.5713461684425525  0.8684371248521288
  0.5600891357250077  0.4248099350595567  0.1494081673745455
  0.4176609468717416  0.8902988886714240  0.8655715856977295
  0.5790832596045945  0.1169976524001655  0.1423064304290165
  0.4587737968849339  0.5803221337999176  0.3795516121170834
  0.5429183457383644  0.4147099981243230  0.6177934685748809
  0.4136603505073495  0.8794809387115876  0.3932701259430331
  0.5762976490016527  0.1262465110905708  0.6052485498380733
  0.1220845283880934  0.5513672043159019  0.3848650642370330
  0.8699567002239590  0.4326158846333783  0.6004456369460769
  0.9977090746156065  0.0050013420808526  0.3813408440191850
  0.9943839569155983  0.9868652025839566  0.6607595163973201
\end{verbatim}

\begin{verbatim}
Cfg28-P-3-Pb9Cu(PO4)6O
   1.00000000000000
     9.8963846483727647   -0.0000017048610227    0.0000000000000000
    -4.9481908476821728    8.5705213635855682    0.0000000000000000
     0.0000000000000000    0.0000000000000000   14.7331589630449589
   Pb   Cu   P    O
    18     2    12    50
Direct
  0.2614347680975300  0.0000410059650520  0.8683395750067149
  0.7385652319024700  0.9999589690349495  0.1316604399932757
  0.9999589690349495  0.2613937321324826  0.8683395750067149
  0.0000410059650520  0.7386062378675291  0.1316604399932757
  0.7386062378675291  0.7385652319024700  0.8683395750067149
  0.2613937321324826  0.2614347680975300  0.1316604399932757
  0.9934254906614797  0.2229951516111939  0.3811981366656809
  0.0065744863385220  0.7770048633888038  0.6188018933343145
  0.7770048633888038  0.7704303530502798  0.3811981366656809
  0.2229951516111939  0.2295696319497225  0.6188018933343145
  0.2295696319497225  0.0065744863385291  0.3811981366656809
  0.7704303530502798  0.9934254906614726  0.6188018933343145
  0.3333333429999996  0.6666666870000029  0.0069718549956121
  0.6666666269999979  0.3333333129999971  0.9930281690043827
  0.3333333429999996  0.6666666870000029  0.2708949996487107
  0.6666666269999979  0.3333333129999971  0.7291050003512893
  0.3333333429999996  0.6666666870000029  0.7339472467749033
  0.6666666269999979  0.3333333129999971  0.2660527822250955
  0.3333333429999996  0.6666666870000029  0.5043211105192640
  0.6666666269999979  0.3333333129999971  0.4956788894807289
  0.3827393584147600  0.4055678419028723  0.3856344782330225
  0.6172606415852471  0.5944321880971160  0.6143655507669834
  0.5944321880971160  0.9771714865118781  0.3856344782330225
  0.4055678419028723  0.0228284834881194  0.6143655507669834
  0.0228284834881123  0.6172606415852471  0.3856344782330225
  0.9771714865118852  0.3827393584147600  0.6143655507669834
  0.0255676728188377  0.6299452699891930  0.8770404858588492
  0.9744323251811622  0.3700547300108070  0.1229594991411389
  0.3700547300108070  0.3956224048296519  0.8770404858588492
  0.6299452699891930  0.6043775951703410  0.1229594991411389
  0.6043775951703410  0.9744323251811622  0.8770404858588492
  0.3956224048296519  0.0255676728188377  0.1229594991411389
  0.8454183961896931  0.5243179493972789  0.8722075233419631
  0.1545815888103164  0.4756820206027186  0.1277924466580274
  0.4756820206027186  0.3211004467924070  0.8722075233419631
  0.5243179493972789  0.6788995532075930  0.1277924466580274
  0.6788995532075930  0.1545815888103164  0.8722075233419631
  0.3211004467924070  0.8454183961896931  0.1277924466580274
  0.5209455905630023  0.3724461829688224  0.3860144608828335
  0.4790544094369977  0.6275538460311907  0.6139855391171736
  0.6275538460311907  0.1484994065941763  0.3860144608828335
  0.3724461829688224  0.8515006224058226  0.6139855391171736
  0.8515006224058226  0.4790544094369977  0.3860144608828335
  0.1484994065941763  0.5209455905630023  0.6139855391171736
  0.0898480866382698  0.7372457817980234  0.7919475760545325
  0.9101519363617214  0.2627542182019695  0.2080524089454627
  0.2627542182019695  0.3526023118402435  0.7919475760545325
  0.7372457817980234  0.6473977181597661  0.2080524089454627
  0.6473977181597661  0.9101519363617214  0.7919475760545325
  0.3526023118402435  0.0898480866382698  0.2080524089454627
  0.3118438806738482  0.3875917544126253  0.2883256648391281
  0.6881561193261447  0.6124082455873747  0.7116743351608719
  0.6124082455873747  0.9242521262612229  0.2883256648391281
  0.3875917544126253  0.0757478737387842  0.7116743351608719
  0.0757478737387842  0.6881561193261447  0.2883256648391281
  0.9242521262612229  0.3118438806738482  0.7116743351608719
  0.2493250676162049  0.2826903523055790  0.4474378532747991
  0.7506749323837951  0.7173096476944139  0.5525621167252055
  0.7173096476944139  0.9666347163106295  0.4474378532747991
  0.2826903523055790  0.0333652836893776  0.5525621167252055
  0.0333652836893776  0.7506749323837951  0.4474378532747991
  0.9666347163106295  0.2493250676162049  0.5525621167252055
  0.0803547255186245  0.7386140815195787  0.9618653497283844
  0.9196452814813867  0.2613859184804213  0.0381346572716197
  0.2613859184804284  0.3417406369990417  0.9618653497283844
  0.7386140815195859  0.6582593630009583  0.0381346572716197
  0.6582593630009583  0.9196452814813867  0.9618653497283844
  0.3417406369990417  0.0803547255186174  0.0381346572716197
  0.1043141586749599  0.5257440936492159  0.8821986583730350
  0.8956858263250496  0.4742558773507710  0.1178013266269460
  0.4742558773507710  0.5785700210257545  0.8821986583730350
  0.5257440936492159  0.4214299789742526  0.1178013266269460
  0.4214299789742526  0.8956858263250425  0.8821986583730350
  0.5785700210257545  0.1043141586749599  0.1178013266269460
  0.4381320515321434  0.5761753664694069  0.4172030736033250
  0.5618679184678470  0.4238246335305860  0.5827968973966691
  0.4238246335305860  0.8619567150627390  0.4172030736033250
  0.5761753664694069  0.1380433149372706  0.5827968973966691
  0.1380433149372706  0.5618679184678470  0.4172030736033250
  0.8619567150627390  0.4381320515321434  0.5827968973966691
  0.0000000000000000  0.0000000000000000  0.3530358680182601
  0.0000000000000000  0.0000000000000000  0.6469641619817352
\end{verbatim}
